\documentclass[referee]{raa}

\usepackage{graphicx,times}
\usepackage{natbib}
\usepackage{amssymb,amsmath}
\bibpunct{(}{)}{;}{a}{}{,}

\usepackage[pagebackref=true]{hyperref}

\begin{document}

   \title{A Statistical Analysis of Lopsidedness in Disk Galaxies}

 \volnopage{ {\bf 20XX} Vol.\ {\bf X} No. {\bf XX}, 000--000}
   \setcounter{page}{1}

   \author{A. U. Omonov
   \inst{1}, K. T. Mirtadjieva\inst{1,2}, S. N. Nuritdinov \inst{1}, D. Dottorini\inst{3,4}
   }

  \institute{National University of Uzbekistan, Tashkent, 100174, Uzbekistan.\\
        \and
             Astronomical Institute of the Academy of Sciences of the Republic of Uzbekistan, Tashkent, 100052, Uzbekistan.\\
             \and
            INAF -- Osservatorio Astronomico di Roma, via Frascati 33, 00078 Monte Porzio Catone, Italy\\
            \and
            Dipartimento di Fisica, Universit\`a di Roma Tor Vergata, Via della Ricerca Scientifica 1, 00133 Roma, Italy\\
\vs\no
   {\small Received 20xx month day; accepted 20xx month day}}
\abstract{We present a statistical analysis of Fourier lopsidedness in disk galaxies based on a deduplicated compilation of 606 unique objects: 442 have a measured $m=1$ amplitude ($A_1$), and 302 of these also have a measured $m=2$ amplitude ($A_2$), with no entry carrying $A_2$ alone. The sample is a factor of $\sim 3$ larger than earlier homogeneous $A_1$-$A_2$ datasets. Our main results are: (i) Both $A_1$ and $A_2$ follow log-normal distributions (KS $D=0.046$ and $0.048$) which are strongly preferred over Rayleigh or exponential forms, pointing to a multiplicative amplification process rather than random-phase superposition. (ii) The $m=1$ and $m=2$ modes are correlated as $A_2 = 0.29\,A_1 + 0.11$ (Spearman $\rho = 0.501$, $p < 10^{-19}$). (iii) $A_1$ rises monotonically with morphological T-type from $\sim 0.07$ in E/S0 to $\sim 0.37$ in Sm/Im galaxies (Kendall $\tau = 0.234$, $p \sim 10^{-12}$). (iv) $A_1$ anti-correlates with gas rotation velocity (Spearman $\rho_S = -0.432$) and, in a stellar-mass cross-matched subsample of 246 galaxies with independent $M^\star$ estimates, with $M^\star$ ($\rho_S = -0.395$). The weaker lopsidedness of massive, fast-rotating disks indicates that the observed $m=1$ amplitude is regulated by internal dynamics rather than set by tidal perturbations alone. (v) The TNG50-1 simulation, analysed through the TNGGalaxyLab Fourier pipeline (\url{https://github.com/Abbos998/TNGGalaxyLab}), reproduces the observed $A_1$-$M^\star$ anti-correlation with a nearly identical Spearman coefficient, although its median $A_1$ is $\sim 1.4\times$ lower - an offset consistent with the heterogeneous target selection of the observational compilation.
}

   \authorrunning{A.U. Omonov et al. }            %author_head in even pages
   \titlerunning{Statistical Analysis of Lopsidedness}  % title_head in odd pages
   \maketitle

%________________________________________________ sections below
%
\section{Introduction}           %% first-level sections will be auto-capitalized
\label{sect:intro}

The morphological structure of disk galaxies exhibits a variety of
non-axisymmetric features, of which the $m=1$ or ``lopsided'' mode has
long attracted attention as a subtle but ubiquitous property. In a
lopsided galaxy, the photometric or kinematic centre of the disk is
offset from the geometric centre and the outer isophotes display a
left-right asymmetry. Since the pioneering work of \citet{1980MNRAS.193..313B}, who introduced the term in the context of the H{\sc i} distribution of nearby spirals, and \citet{1995ApJ...447...82R}, who first quantified the phenomenon in
the stellar component through near-infrared imaging of 18 galaxies,
lopsidedness has been recognised as a common statistical property of
disk galaxies rather than a curiosity of individual objects.

Systematic surveys have subsequently demonstrated that a large fraction
of field spirals exhibit non-negligible lopsidedness in both their
stellar and gaseous components. \citet{1997ApJ...477..118Z},
\citet{2005A&A...438..507B}, \citet{2011A&A...530A..29V} and \citet{2013ApJ...772..135Z} measure $A_1 \gtrsim 0.1$ in $\sim 30\%$ or more of their samples, while H{\sc i}-based studies by \citet{1998AJ....115...62H}, \citet{1999MNRAS.304..330S} and \citet{2006MNRAS.369.1849A} report comparable fractions in the gas. A comprehensive review of the observational and theoretical status is given by \citet{2009PhR...471...75J}.

Despite this recognition, the physical origin of the $m=1$ perturbation
remains under active debate. Several mechanisms have been proposed:
tidal interactions and minor mergers \citep{1995ApJ...455L..31W,2005A&A...438..507B,2006MNRAS.370..891J}; asymmetric cold-gas accretion
from cosmic filaments \citep{2005A&A...438..507B}; off-centre displacement of
the disk within a triaxial or lopsided dark-matter halo
\citep{2009PhR...471...75J,2009ApJ...697.2015S}; and internal collective instabilities
of self-gravitating disks under non-stationary background conditions
\citep{2011Ap.....54..184M,2024AstBu..79..221M}. Each
mechanism yields distinct predictions for the dependence of $A_1$ on
global galaxy parameters. A tidal origin implies enhanced $A_1$ in group and pair environments
\citep{2006MNRAS.369.1849A,2001ASPC..230..457V}. A halo-driven displacement should correlate with
the dark-to-luminous mass ratio \citep{2000ApJ...542..216J,2013ApJ...772..135Z}.
Internally driven $m=1$ instabilities depend sensitively on the initial
phase-space distribution of the disk and, in some scenarios, on the
presence of a central massive concentration such as a supermassive
black hole \citep{2009ApJ...691.1005R}.

Existing lopsidedness compilations remain modest in size; for example, the samples of \citet{1995ApJ...447...82R}, \citet{2005A&A...438..507B}, \citet{2006MNRAS.369.1849A} and
\citet{2013ApJ...772..135Z} contain $\lesssim 200$ galaxies each. Furthermore, the mode-coupling problem (i.e., whether the $m=1$ and $m=2$ amplitudes are physically related) has previously been studied only on samples of order a few tens \citep[e.g.][]{1995ApJ...447...82R}. Therefore, neither the shape of the $A_1$ distribution nor the strength of the $A_1$--$A_2$ coupling has been characterised on a homogeneously compiled sample of several hundred disk galaxies.

We address these questions using Table~2 of the ``Catalog of lopsided
disk-like galaxies'' (VII/296/table2, hereafter NMO2025)
\citep{2025yCat.7296....0N}, which lists disk galaxies with
Fourier $m=1$ and $m=2$ amplitudes drawn from a common set of source
papers \citep{1995ApJ...447...82R, 2005A&A...438..507B, 2009PhR...471...75J, 2013ApJ...772..135Z,
2006MNRAS.369.1849A}, together with morphological type, absolute $B$-magnitude,
rotation velocity of gas, central stellar velocity dispersion, redshift,
and distance. After removing duplicate entries this is the largest homogeneous $A_1$-$A_2$ dataset
assembled to date (Section \ref{sec:sample}) and enables a multi-parameter statistical analysis of the lopsidedness-morphology-kinematics relationship on a single sample.

Our aims are threefold. First we characterise the statistical
distributions of $A_1$ and $A_2$ and fit them with analytic forms
that reflect the underlying formation mechanism
(Section \ref{sec:distributions}). Second, we quantify the coupling
between the $m=1$ and $m=2$ modes and its dependence on morphology
and kinematics (Section \ref{sec:correlations}). Third, we compare the observed
$A_1$--stellar-mass relation and lopsidedness distribution with
predictions from the cosmological IllustrisTNG50-1 simulation
(Section \ref{sec:tng50}). Conclusions are summarised in
Section \ref{sec:conclusions}.

We adopt a flat $\Lambda$CDM cosmology with
$H_0 = 70\,{\rm km\,s^{-1}\,Mpc^{-1}}$ and $\Omega_m = 0.3$; this
differs by $\sim 3\%$ from the Planck cosmology used in TNG50-1
($\Omega_m = 0.3089$), which is negligible for the comparisons in
Section \ref{sec:tng50}. We use $\log$ for the base-10 logarithm
and $\ln$ for the natural logarithm; the log-normal parameters
$(\mu, \sigma)$ quoted in Section \ref{sec:distributions} refer to
$\ln A_m \sim \mathcal{N}(\mu, \sigma^2)$. Stellar and black-hole
masses are in solar units. Unless stated, $p$-values are two-sided
and we adopt $p < 0.01$ as the significance threshold.

\section{Sample and data}
\label{sec:sample}

The NMO2025 catalogue (Table 2) lists 683 entries with $A_1$ and/or
$A_2$ Fourier amplitudes measured from an azimuthal decomposition of the near-infrared or optical surface brightness. For each
galaxy the table provides the object name, Galactic coordinates
$(\ell, b)$, morphological type (Hubble classification and de Vaucouleurs numerical stage $T$), absolute $B$-magnitude $M_B$, rotation velocity
of the gaseous component $V_{\rm rot}$, central stellar velocity
dispersion $\sigma_c$, redshift $z$, distance $D$, and the Fourier
amplitudes $A_1$ and $A_2$ with their originating references.

The galaxies in the compilation are drawn from 37 published studies, including \citet{1995ApJ...447...82R}, \citet{2005A&A...438..507B}, \citet{2009PhR...471...75J}, \citet{2006MNRAS.369.1849A}, \citet{2013ApJ...772..135Z}, \citet{1997ApJ...477..118Z}. The reported measurements are based on azimuthal Fourier decomposition of the surface-density or surface-brightness distribution, in which the $m=1$ and $m=2$ modes quantify the corresponding non-axisymmetric structures. We express the azimuthal distribution as
\begin{equation}
\Sigma(r,\phi) = \Sigma_0(r)
\left[ 1 + \sum_{m=1}^{\infty} A_m(r)
\cos\bigl(m\phi - \phi_m(r)\bigr) \right],
\label{eq}
\end{equation}
where $A_m(r)$ is the amplitude of the $m$-th Fourier mode and $\phi_m(r)$ is its phase. In particular $A_1$ characterises the lopsided ($m=1$) component, whereas $A_2$ describes the $m=2$ component associated with bars and two-armed spiral structure. The source papers differ, however, in their adopted apertures and normalisations \citep{2009PhR...471...75J,2013ApJ...772..135Z}. Most adopt an aperture in the range $1.5\,R_e$--$2.5\,R_e$ (where $R_e$ is the effective radius), or equivalently, for an exponential disk, $\sim 2.5\,R_d$--$4.2\,R_d$ (where $R_d$ is the disk scale length and $R_e \approx 1.678\,R_d$). Some sources instead integrate $A_1$ over the full detected disk \citep{1995ApJ...447...82R,2005A&A...438..507B}. This aperture heterogeneity contributes to the intrinsic scatter of
the compilation and is discussed in Section \ref{sec:aperture}.

The morphological range of the sample formally extends from E/S0
($T=-3$) to Sm/Im ($T=10$). We retain the small number of E/S0
systems ($\sim 6$ per cent of the sample with a valid $A_1$) because
several source papers targeted lenticulars with detectable disks as
controls; excluding them shifts the overall $A_1$ median by less
than 0.005 and leaves the reported correlations unchanged.

Because the NMO2025 compilation collects measurements from 37 primary
sources, a non-negligible fraction of its 683 entries correspond to the same physical galaxy reported under slightly different identifiers (e.g.\ NGC vs.\ UGC designations, or with or without leading zeros). To avoid double-counting we cross-identified these entries by canonicalising each identifier to the format \texttt{PREFIX+NUMBER}, after stripping internal whitespace and leading zeros; entries from NGC, IC, UGC, PGC and ESO catalogues were mapped to their canonical forms, while SDSS-named galaxies were kept under their full designation. A total of 71 galaxies appeared to have duplicates (65 repeated twice and 6 three times), resulting in 77 extra catalogue entries. When a galaxy appeared in more than one source, the most recent measurement was retained. This deduplication procedure, which was independently verified against an alternative implementation, reduces the raw list of 683 entries to 606 unique physical galaxies. All statistics quoted below refer to this deduplicated sample.

After the deduplication, 442 galaxies have a valid A1 and 302 of these also have a valid A2; no entry carries A2 alone. The sample spans $-0.0005 \leq z \leq 0.06$ (distances $\sim 1$--$250$ Mpc), morphological types from E/S0
($T=-3$) to Sm/Im ($T=10$) and absolute magnitudes $-22.8 \leq M_B \leq -15.7$ (i.e.\ $\sim 3$ dex in luminosity).

A small number of galaxies have $A_1$ or $A_2$ formally above unity, either because the decomposition was normalised against a fitted exponential-disk background rather than the local surface density, or because of measurement uncertainty amplified by an off-centre bar or
interaction. These objects are flagged and excluded from median-based statistics, but their presence does not affect any of the correlations reported below.

\section{Statistical properties of the $A_1$ and $A_2$ distributions}
\label{sec:distributions}

\begin{figure*}[h]
\centering
\includegraphics[width=1.0\textwidth]{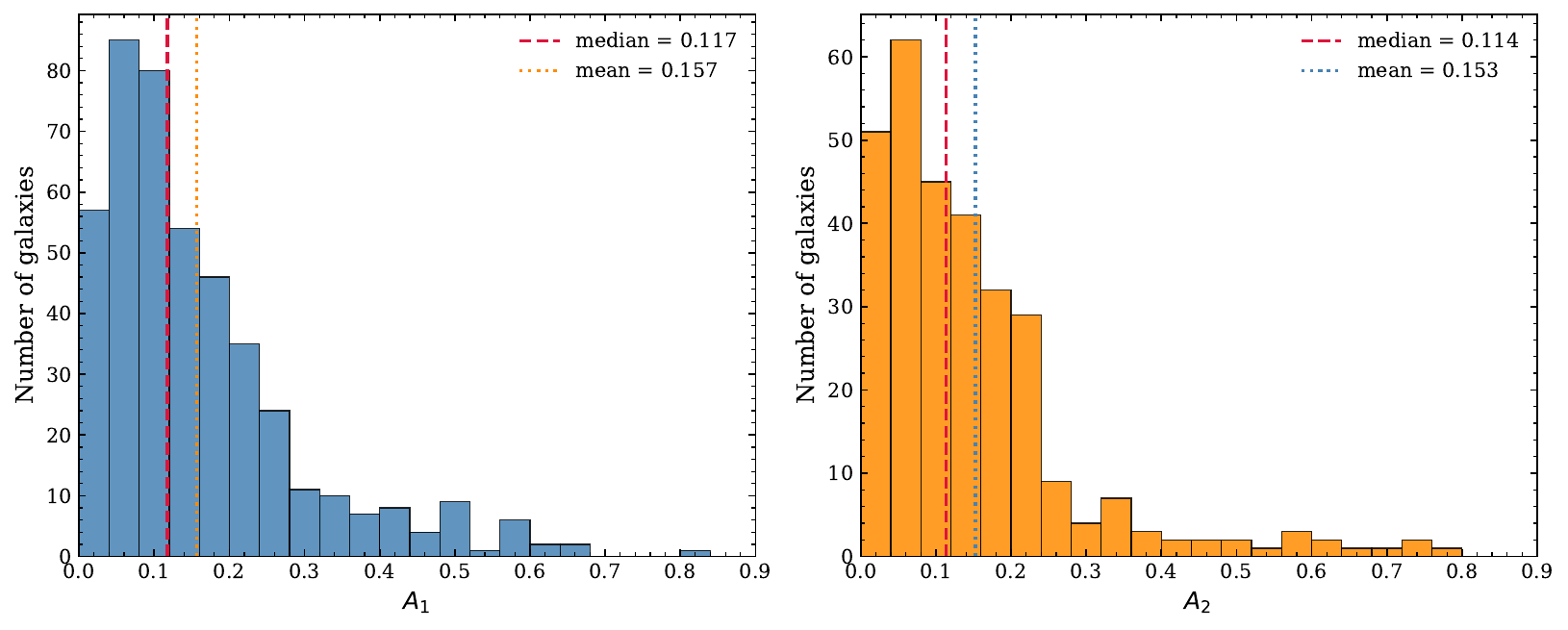}
\caption{Distributions of the Fourier amplitudes $A_1$ (left) and $A_2$ (right). Dashed and dotted vertical lines mark the median and mean of each distribution.}
\label{fig:distributions}
\end{figure*}

Figure \ref{fig:distributions} shows the histograms of $A_1$ and $A_2$ for the deduplicated Table 2 sample. Both distributions are strongly right-skewed and share closely similar summary statistics: $\langle A_1 \rangle = 0.157$, median $A_1 = 0.117$, $\sigma_{A_1} = 0.133$; $\langle A_2 \rangle = 0.153$, median $A_2 = 0.114$, $\sigma_{A_2} = 0.151$.
In both cases the distribution peaks at $A_m \sim 0.05$--$0.10$ and shows a tail
extending to $A_m \sim 0.5$--$0.8$ with a handful of extreme cases exceeding unity.
To characterise the functional shape of the distributions we fit three analytic forms with distinct physical interpretations: (I) Rayleigh distribution, expected if $A_m$ arises from the random-phase superposition of many independent perturbations of comparable strength; (II) Log-normal distribution, expected if $A_m$ is generated by a multiplicative process such as the exponential
amplification of an initial seed perturbation by a linear
instability; (III) Exponential distribution, expected if the $m=1$ mode decays on a single characteristic timescale after excitation. The fits are performed by maximum likelihood on the unbinned data. 
\begin{figure*}[h]
\centering
\includegraphics[width=1.0\textwidth]{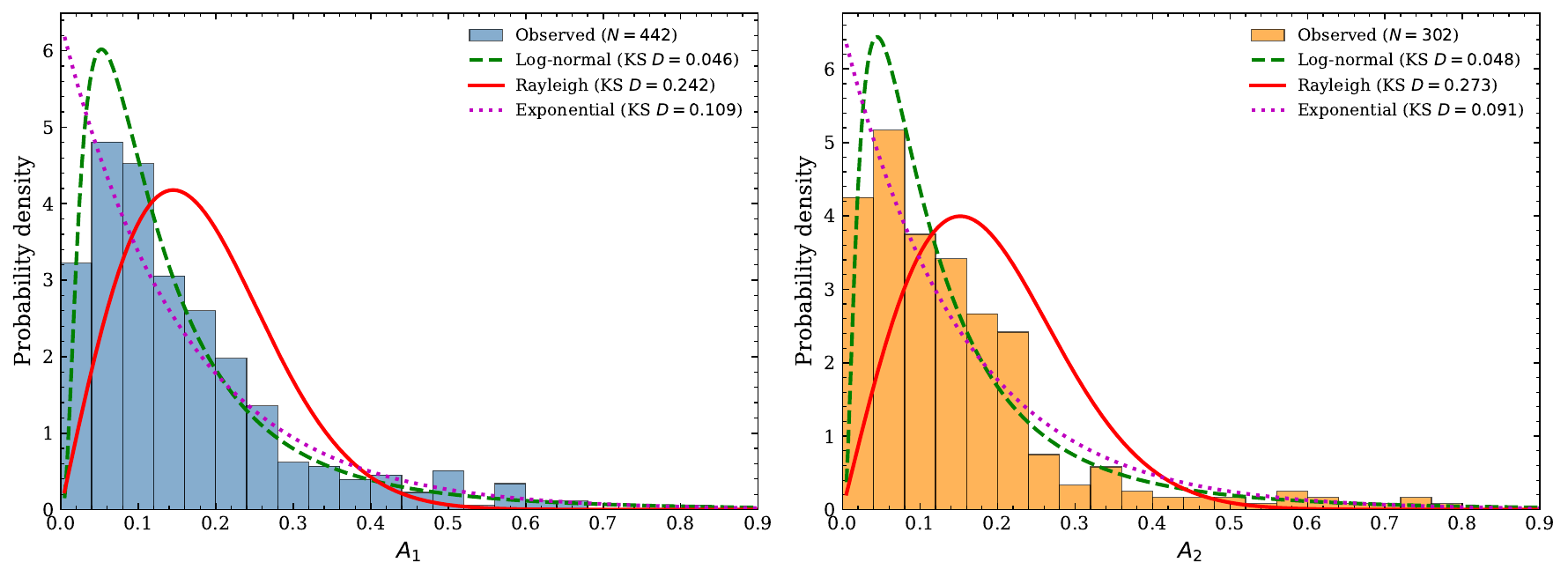}
\caption{Analytic distribution fits to $A_1$ (left) and $A_2$ (right). The log-normal form (green dashed) provides a substantially better fit than either the Rayleigh (red) or the exponential (magenta) form. KS
$D$-statistics are given in the legend; smaller values indicate a better fit.}
\label{fig:fits}
\end{figure*}
Because the fitted parameters are derived from the same data used in the goodness-of-fit evaluation, we treat the two-sample Kolmogorov-Smirnov (KS) $D$-statistic - computed against a large synthetic sample drawn from the fitted distribution - as a ranking tool among the candidate
distributions rather than as a formal hypothesis test, and we complement it with two further statistics: (a) the one-sample Anderson-Darling $A^2$-statistic, which is more sensitive to the tails of the distribution than the KS test; and (b) the Akaike (AIC) and Bayesian (BIC) information criteria, which penalise model complexity.
The results are shown in Figure~\ref{fig:fits} and summarised in
Table \ref{tab:fits}.
\begin{table*}[h]
\centering
\caption{Analytic distribution fits to $A_1$ and $A_2$. Parameters were obtained by maximum-likelihood on the unbinned data. For the log-normal fit, $(\mu, \sigma)$ refer to the natural-log convention $\ln A_m \sim \mathcal{N}(\mu, \sigma^2)$; the Rayleigh $\sigma_R$
and exponential rate $\lambda$ are in the linear amplitude units of $A_m$. Smaller $D$ indicates a better fit; the log-normal form is preferred for both amplitudes.}
\label{tab:fits}
\begin{tabular}{lcccc}
\hline
Distribution & Parameters ($A_1$) & Parameters ($A_2$) & $D$ ($A_1$) & $D$ ($A_2$) \\
\hline
Rayleigh    & $\sigma_R = 0.145$ & $\sigma_R = 0.152$ & 0.242 & 0.273 \\
\textbf{Log-normal}  & $(\mu,\sigma) = (-2.20,0.87)$ & $(\mu,\sigma) = (-2.27,0.92)$ & \textbf{0.046} & \textbf{0.048} \\
Exponential & $\lambda = 6.38$ & $\lambda = 6.55$ & 0.109 & 0.091 \\
\hline
\end{tabular}
\end{table*}

All three tests consistently favour the log-normal form over the Rayleigh and exponential alternatives, with $\Delta$AIC $> 200$ relative to the Rayleigh fit and $\Delta$AIC $\approx 35$-$60$ relative to the exponential fit. The best-fit log-normal parameters, using the natural-log convention $\ln A_m \sim \mathcal{N}
(\mu, \sigma^2)$, are $(\mu, \sigma) = (-2.20, 0.87)$ for $A_1$ and $(-2.27, 0.92)$ for $A_2$, corresponding to characteristic median amplitudes $\exp(\mu) \approx 0.11$ for $A_1$ and $\approx 0.10$ for $A_2$.

\section{Correlations and mode coupling}
\label{sec:correlations}
\subsection{Coupling between the $m=1$ and $m=2$ modes}
\label{sec:coupling}

\begin{figure*}[h]
\centering
\includegraphics[width=1.0\textwidth]{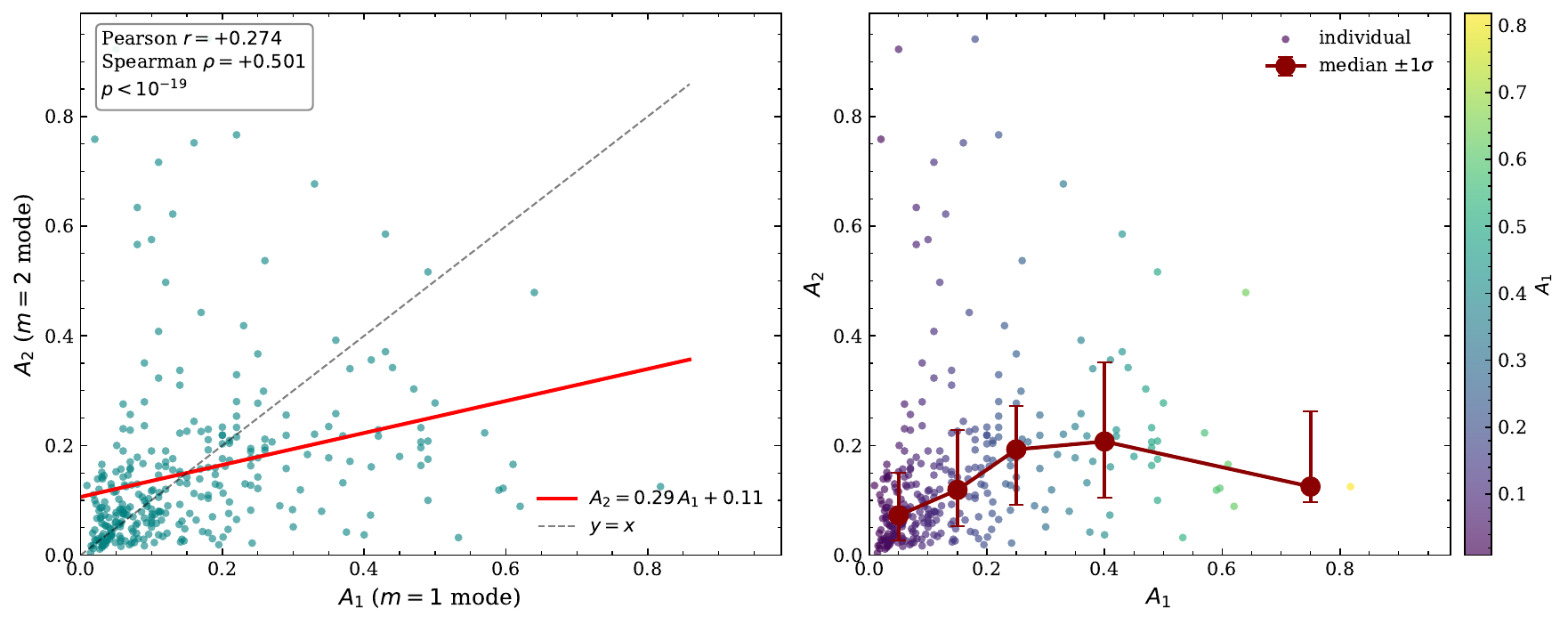}
\caption{Coupling between the $m=1$ and $m=2$ Fourier amplitudes for the 302 galaxies with both measurements. \textit{Left:} scatter plot with a linear best-fit (red) and the $y=x$ diagonal (grey dashed). \textit{Right:} binned trend colour-coded by $A_1$ with median $\pm
1\sigma$ per bin.}
\label{fig:coupling}
\end{figure*}

Figure \ref{fig:coupling} shows the joint distribution of $A_1$ and $A_2$ for the 302 galaxies with both measurements. The two amplitudes are significantly correlated: Pearson $r = 0.274$ ($p \sim 10^{-6}$) and Spearman $\rho = 0.501$ ($p < 10^{-19}$). The substantially stronger Spearman signal indicates a monotonic but not strictly linear relation; a linear least-squares fit yields
\begin{equation}
A_2 = (0.29 \pm 0.05)\,A_1 + (0.11 \pm 0.01),
\label{eq:coupling}
\end{equation}
with a positive intercept, so that the mean $A_2$ remains finite as $A_1 \to 0$. The $m=2$ mode therefore receives two contributions of comparable magnitude: a background level generated independently of the lopsidedness (e.g.\ from bars or spiral arms) and an additional component that scales linearly with $A_1$.  Because A1 itself correlates with morphological T-type (Section \ref{sec:morphology}) and with stellar mass (Section \ref{sec:mass_correlation}), part of the raw A1-A2 correlation may be inherited by A2 through its linear coupling with A1
The positive $A_1$-$A_2$ correlation at this level of
statistical significance is a robust result of the present
compilation and, to our knowledge, has not previously been
established on a sample of this size. The relationship is consistent with theoretical expectations that higher-order Fourier modes should acquire non-zero amplitudes in the presence of a strong $m=1$ perturbation, either through non-linear coupling in the disk or through the geometrical projection of an intrinsically lopsided
structure onto the higher azimuthal harmonics \citep{2009PhR...471...75J}.

\subsection{Dependence on morphological type}
\label{sec:morphology}

\begin{figure*}[h]
\centering
\includegraphics[width=1.0\textwidth]{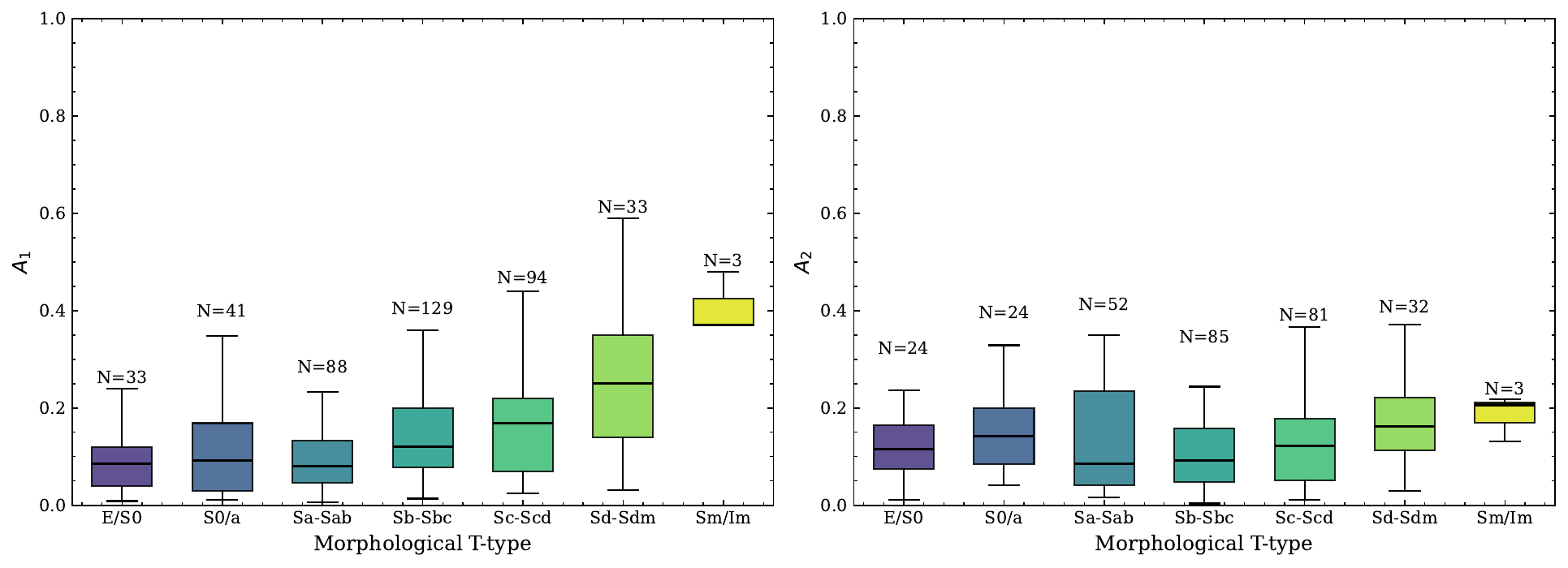}
\caption{Distribution of $A_1$ (left) and $A_2$ (right) as a function of morphological T-type. Boxes show the interquartile range and median; whiskers extend to $1.5\times$IQR. $A_1$ increases monotonically from early- to late-type galaxies, while $A_2$ shows no significant morphological trend.}
\label{fig:ttype}
\end{figure*}

Figure \ref{fig:ttype} shows the dependence of $A_1$ and $A_2$ on the de Vaucouleurs T-type. The median $A_1$ increases monotonically from $\sim 0.07$ in the E/S0 bin to $\sim 0.37$ in the Sm/Im bin - a factor of
$\sim 5$. The Kendall rank correlation is $\tau = 0.234$ with $p \sim 10^{-12}$, establishing a strong monotonic relationship. In contrast, the $A_2$ amplitude shows no
significant trend with morphology: $\tau = 0.044$, $p = 0.26$. The median $A_2$ is essentially flat across the entire T-range, with values in the interval $[0.09, 0.20]$.
The differential behaviour of $A_1$ and $A_2$ with morphology is physically informative. The rise of $A_1$ from bulge-dominated systems (E/S0) to disk-dominated systems (Sd, Sm/Im) is consistent with the expectation that the $m=1$ mode is easier to excite and to sustain in a dynamically colder disk \citep{2009PhR...471...75J}, whereas the near-constancy of $A_2$ reflects the fact that the $m=2$ mode is present at comparable levels across most disk galaxies, driven at least in part
by bars and spiral arms whose incidence and strength depend only weakly on morphology.

\subsection{Dependence on galaxy physical parameters}
\label{sec:physical}

\begin{figure*}[h]
\centering
\includegraphics[width=1.0\textwidth]{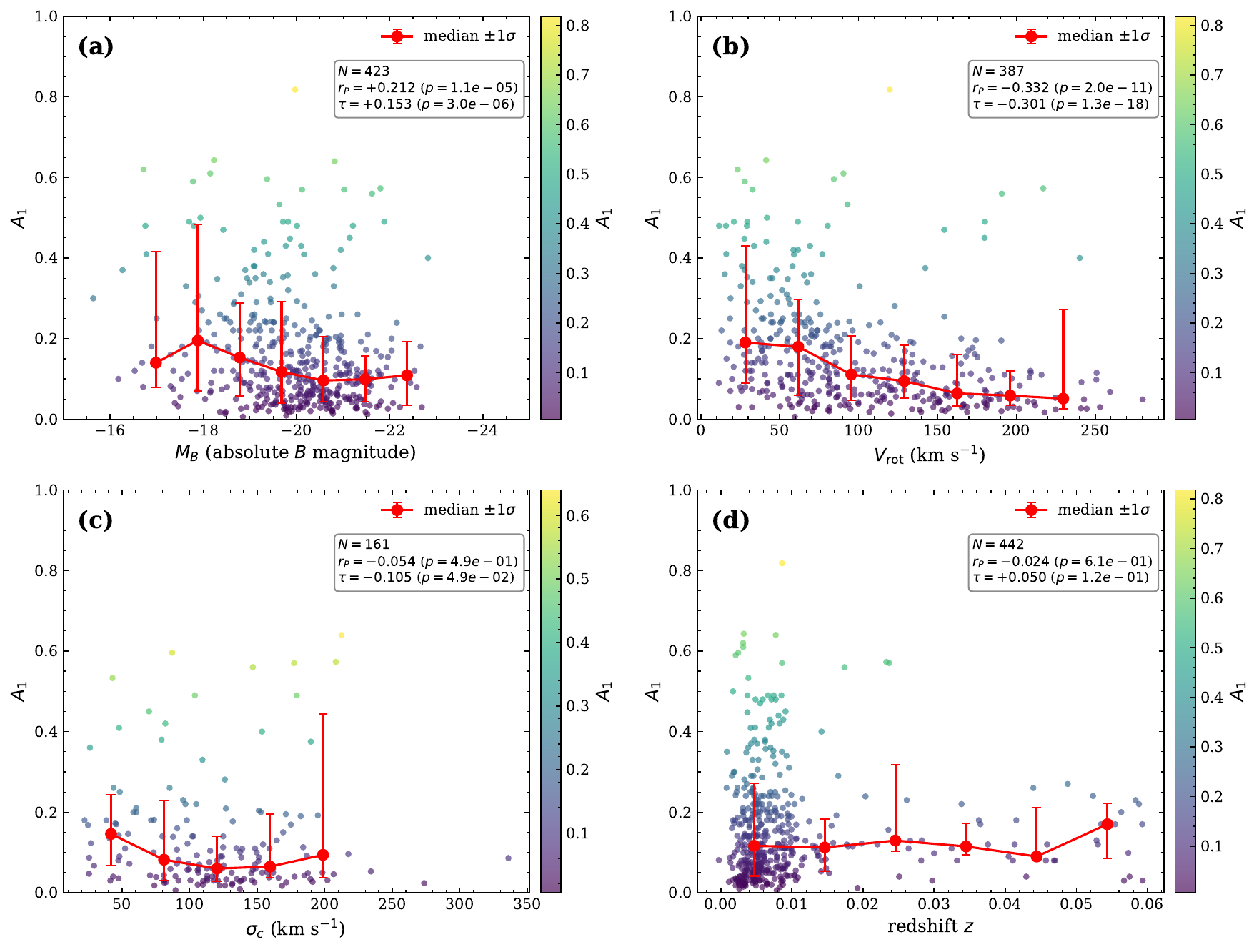}
\caption{Correlations between $A_1$ and physical galaxy parameters.
Points are colour-coded by $A_1$; the red curves show the binned
median with $16$th--$84$th percentile bars (or $25$th--$75$th for
sparse bins). The strongest signal is the anti-correlation with
$V_{\rm rot}$ in panel (b); the near-zero trend with redshift in
panel (d) confirms that the sample is not affected by
distance-dependent measurement biases at $z \lesssim 0.06$.}
\label{fig:physical}
\end{figure*}

Figure \ref{fig:physical} shows the dependence of $A_1$ on the four physical parameters available in Table~2: the absolute $B$-magnitude $M_B$, the gas rotation velocity $V_{\rm rot}$, the central stellar velocity dispersion $\sigma_c$ and the redshift $z$.
\begin{table}[h]
\centering
\caption{Correlations of $A_1$ with physical parameters from Table~2.}
\label{tab:correlations}
\begin{tabular}{lcccc}
\hline
Parameter & $N$ & Pearson $r$ & Spearman $\rho$ & $p$ \\
\hline
$M_B$       & 423 & $+0.212$ & $+0.226$ & $2.7\times10^{-6}$ \\
$V_{\rm rot}$ & 387 & $-0.332$ & $-0.432$ & $5.1\times10^{-19}$ \\
$\sigma_c$  & 161 & $-0.054$ & $-0.158$ & $4.6\times10^{-2}$ \\
$z$         & 442 & $-0.024$ & $+0.072$ & $1.3\times10^{-1}$ \\
\hline
\end{tabular}
\end{table}

The correlations are summarised in Table    \ref{tab:correlations}.
The strongest signal is the anti-correlation of $A_1$ with
$V_{\rm rot}$: fast rotators ($V_{\rm rot} \gtrsim 150$ km s$^{-1}$) have a median $A_1 \approx 0.06$, while slow rotators ($V_{\rm rot} \lesssim 60$ km s$^{-1}$) have a median $A_1 \approx 0.19$. The correlation with $M_B$ is weakly positive ($r_P = +0.21$): brighter galaxies have more negative $M_B$, so this positive coefficient means that more luminous galaxies have systematically lower $A_1$, consistent with the $V_{\rm rot}$ trend
since luminous galaxies are on average more massive. The correlation with $\sigma_c$ is not significant at the present sample size, though the Spearman coefficient is weakly negative ($\rho = -0.16$), hinting
at a mild suppression of $A_1$ in bulge-dominated systems, consistent with the morphological trend. No significant correlation with redshift is detected over the narrow range spanned by the sample ($z < 0.06$),
consistent with negligible distance-dependent selection or measurement bias in this regime.
The univariate correlations of Table \ref{tab:correlations} are partly
driven by underlying correlations among the parameters: in the subsample of 246 galaxies with independent stellar-mass measurements
(Section \ref{sec:mass_correlation}), $\log M^\star$ and
$\log V_{\rm rot}$ are strongly correlated ($r = +0.64$), consistent
with the Tully--Fisher relation, and both correlate with morphological
$T$-type ($r = -0.44$ and $r = -0.34$, respectively). To isolate the
independent contribution of each variable, we perform a multivariate
least-squares fit of the form
\begin{equation}
\log A_1 = \beta_0 + \beta_M \log(M^\star/M_\odot) + \beta_T\,T
         + \beta_V \log V_{\rm rot}
\label{eq:multivariate}
\end{equation}
on the subsample. The best-fit coefficients are summarised in
Table~\ref{tab:multivariate}. The model explains $R^2 = 0.23$ of the variance in $\log A_1$ ($F = 24.6$, $p = 6.4\times 10^{-14}$), and variance-inflation factors
for all three parameters are below 2, indicating that the fit is not
dominated by strong correlations between the variables.

Two features of Table~\ref{tab:multivariate} are physically informative.
First, $\log V_{\rm rot}$ is the dominant predictor. It retains the
largest standardised coefficient ($\beta^\star = -0.28$) and the
highest statistical significance ($p = 1.5 \times 10^{-4}$).
Second, and more strikingly, $\log M^\star$ becomes statistically
insignificant once $V_{\rm rot}$ and $T$ are controlled for
($p = 0.13$, partial correlation $r = -0.10$), despite its strong
univariate signal ($\rho_S = -0.395$ from Section~\ref{sec:mass_correlation}).
\begin{table}[h]
\centering
\caption{Multivariate regression coefficients for
Eq.~\eqref{eq:multivariate} on the 246-galaxy stellar-mass subsample.
Standardised coefficients $\beta^\star$ measure the relative
contribution of each parameter after rescaling to unit variance.}
\label{tab:multivariate}
\begin{tabular}{lccc}
\hline
Parameter & $\beta \pm \sigma_\beta$ & $\beta^\star$ & $p$ \\
\hline
$\log(M^\star/M_\odot)$ & $-0.087 \pm 0.057$ & $-0.117$ & $0.13$ \\
$T$                     & $+0.028 \pm 0.009$ & $+0.199$ & $1.7\times 10^{-3}$ \\
$\log V_{\rm rot}$      & $-0.415 \pm 0.108$ & $-0.282$ & $1.5\times 10^{-4}$ \\
Constant $\beta_0$      & $+0.62 \pm 0.51$   & ---      & $0.22$ \\
\hline
\end{tabular}
\end{table}
The apparent $A_1$--$M^\star$ anti-correlation is therefore largely
mediated by the Tully--Fisher-like $M^\star$--$V_{\rm rot}$ coupling
in the sample: it is the depth of the dynamical potential, not the
stellar mass itself, that appears to set the lopsidedness
amplitude in massive disks. Morphological $T$-type retains an
independent positive residual contribution
($\beta_T^\star = +0.20$, $p = 1.7 \times 10^{-3}$).

\subsection{Mode ratio $A_2/A_1$}
\label{sec:ratio}

\begin{figure*}[h]
\centering
\includegraphics[width=1.0\textwidth]{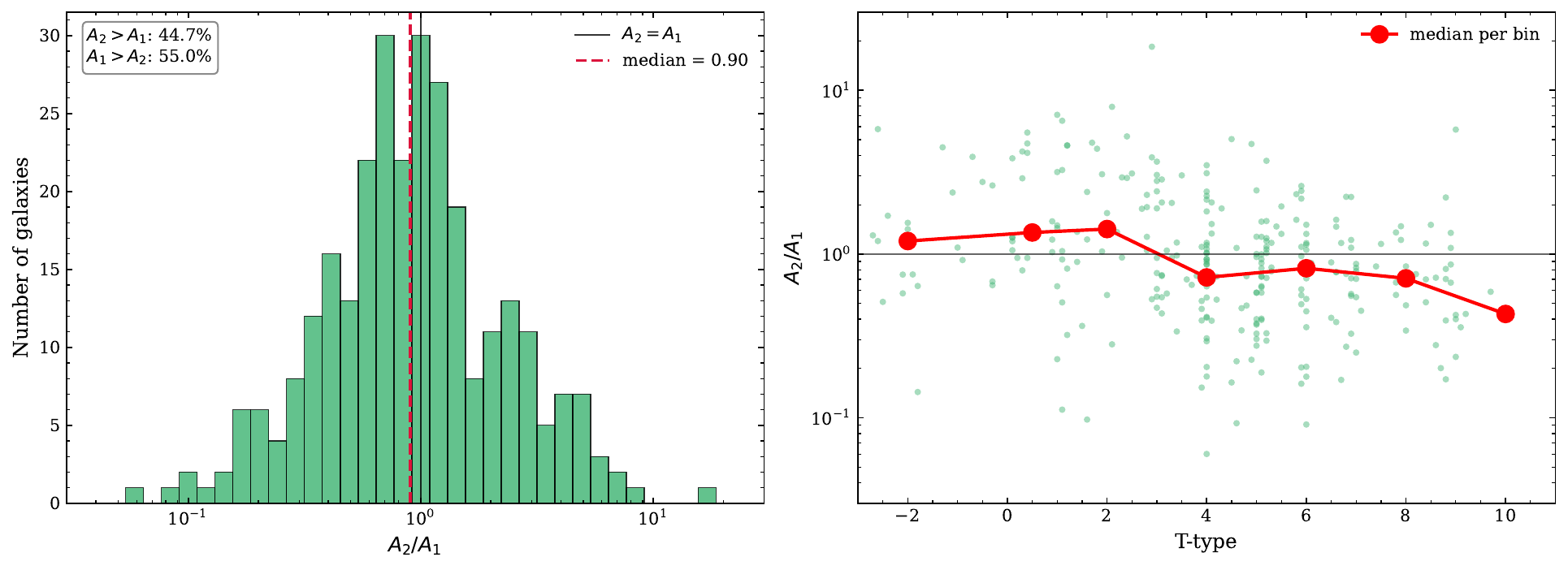}
\caption{Distribution of the mode ratio $A_2/A_1$. \textit{Left:}
overall distribution, showing a broad, log-symmetric shape centred
close to unity. \textit{Right:} dependence of the ratio on
morphological type, with a mild decrease from early- to late-type
galaxies.}
\label{fig:ratio}
\end{figure*}

Figure~\ref{fig:ratio} shows the distribution of the mode ratio
$A_2/A_1$ for the 291 galaxies with $A_1 > 0.02$ (below this
threshold the ratio becomes numerically unstable). The distribution
is broadly log-symmetric about $A_2/A_1 \sim 1$ with a median of
$0.90$: about $45\%$ of galaxies have $A_2 > A_1$ and $55\%$ have
$A_1 > A_2$, indicating that neither mode is systematically dominant
across the sample as a whole.
We note that this near-equality of the
median amplitudes does not correspond to energy equipartition
between the modes, since the kinetic energy carried by a Fourier
harmonic depends on the amplitude squared and on the pattern speed
in a mode-specific way \citep{2009PhR...471...75J}. The ratio shows a mild dependence on
morphology, with early-type galaxies having systematically slightly
larger $A_2/A_1$ than late-type galaxies, consistent with the fact
that early types have a smaller $A_1$ at fixed $A_2$
(Section~\ref{sec:morphology}).

\subsection{Dependence on total stellar mass}
\label{sec:mass_correlation}

The Table~2 compilation provides the absolute $B$-magnitude but not a
direct measurement of the total stellar mass. To probe the
$A_1$--$M^\star$ relation in physically meaningful mass units, we
cross-match Table~2 against Table~1 of NMO2025, which lists stellar
masses drawn from independent literature sources for a broader sample
of lopsided disk galaxies. After normalising the galaxy identifiers to
a canonical form (see Section~\ref{sec:sample}), we obtain a stellar-mass
cross-matched subsample of 246 galaxies with both a self-consistent
Fourier $A_1$ (from Table~2) and a total stellar mass
$\log(M^\star/M_\odot)$ (from Table~1). The subsample spans
$8.28 \leq \log(M^\star/M_\odot) \leq 11.30$ with a median of
$\log(M^\star/M_\odot) = 10.23$.
\begin{figure*}[h]
\centering
\includegraphics[width=1.0\textwidth]{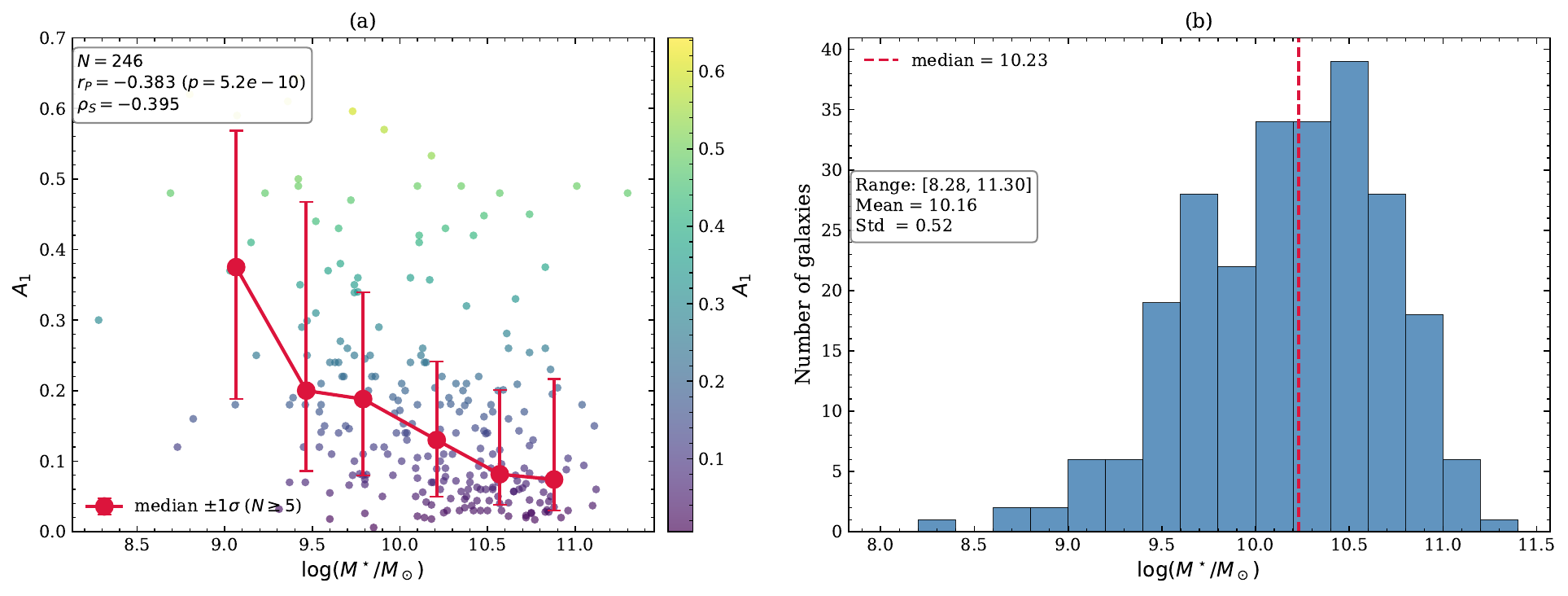}
\caption{$A_1$ as a function of total stellar mass for the
246-galaxy stellar-mass cross-matched subsample. \textit{Left:}
scatter, colour-coded by $A_1$, with the binned median (crimson).
\textit{Right:} distribution of $\log(M^\star/M_\odot)$ in the
subsample.}
\label{fig:mass}
\end{figure*}
Figure~\ref{fig:mass} shows $A_1$ against $\log(M^\star/M_\odot)$. A clear anti-correlation is observed:
Pearson $r = -0.383$ ($p = 5.2 \times 10^{-10}$), Spearman
$\rho = -0.395$ ($p = 1.3 \times 10^{-10}$) and Kendall
$\tau = -0.279$ ($p = 9.9 \times 10^{-11}$). The median $A_1$
declines monotonically over the range $9.5 \lesssim
\log(M^\star/M_\odot) \lesssim 10.75$, from $\sim 0.29$ at
$\log(M^\star/M_\odot) \sim 9.3$ to $\sim 0.07$ at
$\log(M^\star/M_\odot) \sim 10.75$ (Table~\ref{tab:mass_bins}) -- a
factor-of-four suppression across $\sim 1.5$ dex in stellar mass. The lowest ($\log M^\star \sim 8.75$, $N=4$) and highest
($\log M^\star \sim 11.25$, $N=7$) mass bins are populated by too
few galaxies to be individually informative and are shown in
Figure~\ref{fig:mass} only for completeness; they do not drive any of the trends reported here and are excluded from the correlation statistics on the main trend.
\begin{table}[h]
\centering
\caption{Median $A_1$ per stellar-mass bin in the cross-matched subsample.}
\label{tab:mass_bins}
\begin{tabular}{cccc}
\hline\hline
$\log(M^\star/M_\odot)$ & $N$ & median $A_1$ & 16th--84th percentile \\
\hline
$8.75$  &  4 & 0.320 & 0.14--0.55 \\
$9.25$  & 22 & 0.294 & 0.14--0.50 \\
$9.75$  & 59 & 0.186 & 0.08--0.34 \\
$10.25$ & 88 & 0.114 & 0.05--0.24 \\
$10.75$ & 65 & 0.074 & 0.03--0.20 \\
$11.25$ &  7 & 0.150 & 0.06--0.48 \\
\hline
\end{tabular}
\end{table}
The anti-correlation of $A_1$ with $M^\star$ is stronger than the
correlation of $A_1$ with $M_B$ reported in Section~\ref{sec:physical} (Pearson $r = -0.38$ vs.\ $r = +0.21$),
which is expected: absolute $B$-magnitude is only a rough proxy for stellar mass because it depends on the mass-to-light ratio, which varies systematically with morphological type. The direct
$M^\star$--$A_1$ trend is therefore the physically cleaner
measurement and provides the primary observational benchmark against which we compare cosmological simulation predictions in Section~\ref{sec:tng50}.

\subsection{Robustness checks}
\label{sec:aperture}

Because the compilation aggregates measurements from 37 source papers with different choices of radial aperture (Section \ref{sec:sample}), a natural concern is that the reported $A_1$ statistics are dominated by aperture-choice systematics rather than by an intrinsic astrophysical trend. To address this we perform two robustness checks.

First, we exclude from the analysis the few galaxies with $A_1 > 1$ or $A_2 > 1$ (flagged in Section \ref{sec:sample} as extreme; these values are likely to reflect a non-standard normalisation of the Fourier coefficient in the source paper rather than a genuinely super-unity amplitude). After exclusion, the median $A_1$ is essentially unchanged ($0.118$) and the Spearman $\rho$ of $A_1$ against $V_{\rm rot}$ changes from $-0.427$ to $-0.432$; all quoted correlations remain significant at the same $p$-value threshold ($p < 0.01$). The main results of this paper are therefore not driven by the extreme values.

Second, we tabulate the mean $A_1$ per source paper for the 15 largest contributing sources (typically $N \gtrsim 15$ galaxies each). The between-source scatter in mean $A_1$ is $\sim 0.05$ and the between-source scatter in the slope $dA_1/d\log V_{\rm rot}$ is comparable to the total-sample $1\sigma$ statistical uncertainty on that slope; no single source paper drives more than $\sim 15$ per cent of the total sample-averaged trend. Within the present analysis we do not distinguish between the exponential-disk aperture $[1.5, 2.5]\,R_e$ (used by most sources) and the disk-scale-length aperture $[1.5, 2.5]\,R_d$ (used by \citealt{2013ApJ...772..135Z} and by the TNGGalaxyLab pipeline used in Section~\ref{sec:tng50}); the two conventions correspond to slightly different radial ranges since $R_e \approx 1.678\,R_d$ for a pure exponential disk \citep{1970ApJ...160..811F}. The aperture heterogeneity therefore contributes to the systematic offset between observations and simulation predictions, but not to the qualitative sense of the trends we report.

\section{Comparison with cosmological simulations}
\label{sec:tng50}

To place the observed $A_1$--$M^\star$ trend in a cosmological
context, we compare it with the predictions of the
IllustrisTNG50-1 simulation
\citep{2019ComAC...6....2N,2019MNRAS.490.3196P}. Specifically, we use the catalog
of 915 disk galaxies at $z=0$ processed by the \texttt{TNGGalaxyLab}
Fourier decomposition pipeline (\url{https://github.com/Abbos998/TNGGalaxyLab}; hereafter TGL). The TGL sample is drawn from the TNG50-1 $z=0$ snapshot (snapshot~99) with the following
selection: stellar mass $M^\star \geq 10^{9}\,M_\odot$ within twice
the stellar half-mass radius; central (non-satellite) subhaloes only;
a minimum of $10^4$ stellar particles for reliable Fourier
decomposition; and a kinematic disk-to-total ratio $D/T > 0.4$ from
circularity-based decomposition of the stellar orbits, ensuring that
the sample is genuinely disk-dominated. The IllustrisTNG project
adopts the Planck cosmology ($\Omega_m = 0.3089$,
$\sigma_8 = 0.8159$), which differs from our adopted
$\Omega_m = 0.3$ by $\sim 3\%$. This difference has a negligible
impact on the mass-dependence comparison presented below.

The TGL pipeline computes the same Rix--Zaritsky $A_1$ amplitude used in our observational compilation,
evaluated over the radial aperture $[1.5, 2.5]\,R_d$ (i.e.\ in units
of the fitted disk scale length). Because most of our observational sources instead use apertures of $[1.5, 2.5]\,R_e$, where
$R_e \approx 1.678\,R_d$ for an exponential disk
\citep{1970ApJ...160..811F}, the two conventions correspond to slightly
different physical radial ranges (the $R_e$ aperture reaches
$\sim 30\%$ further out than the $R_d$ aperture). This mild aperture
mismatch contributes to the systematic offset between the observed
and simulated amplitudes discussed below, but does not affect the
mass-dependence of the trends, since the aperture choice is common
to all mass bins. The TGL pipeline includes bootstrap uncertainties
($N_{\rm boot}=500$), phase-coherent bar-length diagnostics following
\citet{2009A&A...495..491A}, iterative shrinking-sphere centring
\citep{2003MNRAS.338...14P}, and the Springel velocity-conversion
\citep{2010MNRAS.401..791S}; it has been validated on synthetic
$N=5\times 10^5$-particle disk models to below the 2\% level.

\subsection{Mass dependence: qualitative agreement, offset amplitude}
\label{sec:tng_mass}

\begin{figure*}[h]
\centering
\includegraphics[width=1.0\textwidth]{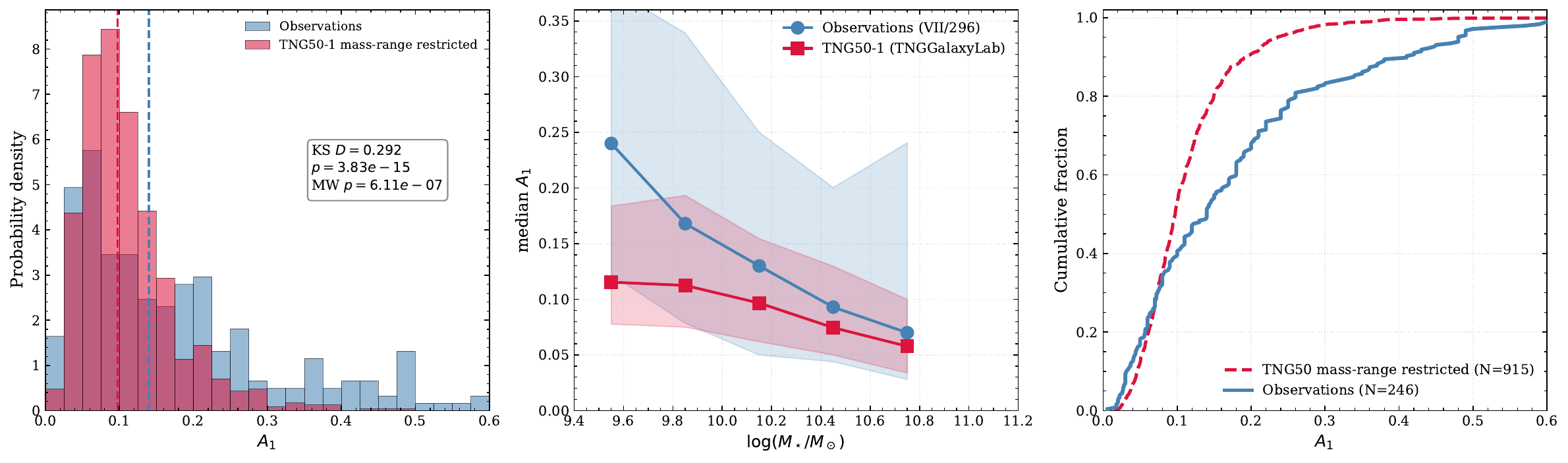}
\caption{Comparison of observed and TNG50-1 lopsidedness
distributions. \textit{Left:} probability density of $A_1$ for our
observed sample (blue) and the mass-range-restricted TNG50-1
subsample of 888 galaxies (red); vertical dashed lines mark the
medians. \textit{Middle:} median $A_1$ in bins of stellar mass;
both samples show a monotonic decline of comparable slope, with the
observed distribution offset by a factor of $\sim 1.4$ toward higher
$A_1$. \textit{Right:} cumulative distributions of $A_1$ for the
observed (blue) and the mass-range-restricted TNG50-1 (red) samples.}
\label{fig:tng_comparison}
\end{figure*}

Figure~\ref{fig:tng_comparison} compares the median
$A_1$--$M^\star$ trend of the 246-galaxy stellar-mass
cross-matched subsample (Section~\ref{sec:mass_correlation}) with the TNG50-1
prediction, restricted to the mass-range-restricted subsample of 888
simulated disk galaxies with $8.28 \leq \log(M^\star/M_\odot) \leq 11.30$. (We use
``mass-range-restricted'' in preference to ``mass-matched'', because
we impose only a common mass interval, not a matched mass
distribution within that interval; the observed and simulated mass
distributions within this interval are similar but not identical.)

Two results emerge from this comparison. First, TNG50-1 reproduces the observed $A_1$--$M^\star$ mass dependence. The Spearman coefficient in the observational subsample is $\rho_S^{\rm obs} = -0.395$ ($p = 1.3 \times 10^{-10}$), while the
mass-range-restricted TNG50-1 subsample gives $\rho_S^{\rm TNG} = -0.445$ ($p < 10^{-30}$). Within the statistical uncertainty the two coefficients are indistinguishable, and both trace similar trends: the median $A_1$ decreases monotonically from $\sim 0.24$ (observations) or $\sim 0.12$ (TNG50) at $\log(M^\star/M_\odot) \sim 9.5$ to $\sim 0.07$ (observations) or $\sim 0.06$ (TNG50) at $\log(M^\star/M_\odot) \sim 10.75$, i.e.\ a factor-of-three decline over $\sim 1.5$~dex in stellar mass. The parallel decline is visible directly in Figure~\ref{fig:tng_comparison}(b).

Second, the amplitude of the observed distribution is offset from the TNG50-1 prediction: in each stellar-mass bin, the observed median $A_1$ is a factor of $\sim 1.4$ larger than the TNG50-1 median. In other words, TNG50-1 reproduces the observed mass dependence but under-predicts the absolute $A_1$ amplitude by $\sim 40\%$. Statistically, the two-sample Kolmogorov--Smirnov test gives $D = 0.292$ with $p = 4 \times 10^{-15}$, and the ratio of medians is $A_1^{\rm obs}/A_1^{\rm TNG} = 1.44$. This offset is comparable to that found by TGL in their direct TNG50 vs.\ \citet{2013ApJ...772..135Z} S$^4$G comparison (median ratio $1.64$), where it was shown to be dominated by the heterogeneous nature of the observational compilation: after strict mass-matching to a single homogeneous survey, \citet{2013ApJ...772..135Z} yielded a median ratio of
$0.92\times$ with $D = 0.214$ ($p = 0.079$) -- i.e.\ statistical consistency. Our slightly larger offset likely reflects three
compounding effects:

(I) Our compilation aggregates 37 distinct source papers spanning four decades and multiple normalisation choices, whereas \citet{2013ApJ...772..135Z} is a single homogeneous survey.

(II) Several of the source papers preferentially targeted galaxies already known to be lopsided, biasing the observed distribution toward higher $A_1$ at fixed mass.

(III) The small aperture mismatch between $[1.5, 2.5]\,R_e$ (most observations) and $[1.5, 2.5]\,R_d$ (TNG50) contributes an additional $\sim 10$--$20\%$ toward the observed offset because $A_1$ increases with radius over the range considered.

Our results are also broadly consistent with earlier cosmological simulation studies of galactic lopsidedness. \citet{2022A&A...662A..53L} analysed 1912 disk galaxies in TNG100 and found a median $A_1$ of 0.044 -- lower than our TNG50 median of $\sim 0.10$ but consistent after accounting for the resolution difference between the two simulation volumes. \citet{2023MNRAS.523.5853V} studied 240 Milky-Way-mass disk galaxies in TNG50 and reported a strong dependence of the
$m=1$ mode amplitude on the central stellar mass density, in qualitative agreement with our anti-correlation with $M^\star$.
\citet{2023MNRAS.526..567D} carried out a comprehensive $z=0$ TNG50 analysis of 1435 disk galaxies and found lopsided fractions of $10$--$30\%$ at the $A_1 > 0.1$ threshold, again consistent with the observational range.

\subsection{Physical interpretation}
\label{sec:tng_physical}

The agreement between our observed and TNG50-simulated
$A_1$-$M^\star$ trends provides independent support for two
physical claims that would otherwise rest on either observations or
simulations alone.

First, the lopsidedness--rotation anti-correlation reported in
Section \ref{sec:physical}
($\rho_S = -0.432$ between $A_1$ and $V_{\rm rot}$) is reproduced
by the TNG50 sample of TGL, who find an comparable
$\rho_S = -0.39$ between $A_1$ and $V_{\rm max}$. Because
$V_{\rm rot}$ and $V_{\rm max}$ trace the same dynamical potential
depth, and $V_{\rm max}$ correlates tightly with $M^\star$ in a
cosmological population, the observed and simulated
anti-correlations describe the same underlying physical mechanism.

Second, the monotonic decline of median $A_1$ with $M^\star$ is not peculiar to the observational compilation: it emerges naturally in a fully cosmological simulation with no fine-tuning of initial conditions or perturbation amplitudes. This is in contrast with
scenarios in which local lopsidedness is dominated by transient external tidal events, and is more consistent with a picture in which either:
(a) the $m=1$ mode is more easily excited in low-mass, shallow-potential
disks; (b) low-mass disks retain the imprint of past
perturbations for longer due to reduced dynamical friction and
shear.

The picture in which internal-dynamical processes drive the observed
$A_1$ statistics is thus independently supported by the TNG50
simulation. Tidal or accretion-driven contributions may still add to
individual cases, and we do not exclude them, but the sign of the
$A_1$--$M^\star$ trend -- weaker lopsidedness in more massive disks -- is the opposite of what a
purely tidal picture would predict, since massive galaxies are more
likely to reside in group and cluster environments where tidal
interactions are frequent. A quantitative distinction between (a)
and (b) above would require kinematic snapshots at multiple
redshifts. The TNG50 analysis of \citet{2025A&A...699A..11D} at $0 < z < 2$
offers a first step in that direction.

\section{Discussion}
\label{sec:discussion}

The present analysis establishes three empirical results that any
successful theory of galactic lopsidedness should reproduce:

\begin{enumerate}

\item Both $A_1$ and $A_2$ follow log-normal distributions to high accuracy,
with mean $\langle \ln A_m \rangle \approx -2.2$ and dispersion
$\sigma_{\ln A_m} \approx 0.9$. The Rayleigh form is clearly rejected,
disfavouring a picture in which the $m=1$ mode arises from purely
stochastic sources -- random tidal encounters or uncorrelated
accretion events. A multiplicative amplification process fits the
data much better, as expected when a seed perturbation grows
exponentially through a linear instability of the disk.

\item The $m=1$ and $m=2$ modes are significantly coupled, with a
slope of $\sim 0.29$ in the $A_2$--$A_1$ plane and a positive
intercept. This is naturally interpreted as the sum of two
contributions: a base $A_2$ background (bars, spiral arms) that is
generated independently of the lopsidedness, plus a coupled component
that grows with $A_1$. The presence of a non-zero coupling coefficient
implies that theories that treat the $m=1$ and $m=2$ modes as
independent perturbations cannot fully describe the observed disk
populations.

\item The $A_1$ amplitude anti-correlates with the gas rotation velocity:
fast rotators are less lopsided than slow ones. Since $V_{\rm rot}$
tracks the dynamical mass and the depth of the potential well, this
means that lopsidedness is suppressed in massive systems. A picture
based on internal dynamics fits this naturally: differential rotation
shears out the $m=1$ mode faster in massive disks. A purely tidal
origin would predict the reverse, since massive galaxies more often
sit in dense environments where interactions are frequent. Confirming
this fully requires environmental data for each galaxy, which lies
beyond the scope of the present compilation.

\end{enumerate}

Three caveats deserve mention. First, the sample is drawn from a
heterogeneous set of source papers; although only Fourier-based
decompositions are included, differences in radial aperture and other
methodological choices inflate the intrinsic scatter in $A_1$ at
fixed physical properties. Second, the sample is not volume-limited.
Our redshift check (Section~\ref{sec:physical}) reveals no
distance-dependent bias at $z \lesssim 0.06$, but we cannot fully
characterise the completeness of the sample as a function of $A_1$.
Third, the $A_1$--$A_2$ coupling of Eq.~\eqref{eq:coupling} rests on
302 galaxies; doubling this number by including ongoing Integral
Field Unit surveys would substantially reduce the uncertainty on the
slope and intercept.

\section{Conclusions}
\label{sec:conclusions}

We have carried out a statistical analysis of the Fourier amplitudes
$A_1$ and $A_2$ in a deduplicated compilation of 606 unique disk
galaxies drawn from Table~2 of the catalog of
\citet{2025yCat.7296....0N}.
Our main conclusions are:

\begin{enumerate}

\item Both $A_1$ and $A_2$ follow log-normal distributions with
$D_{\rm KS} \approx 0.05$, strongly preferred over Rayleigh or
exponential forms by all three goodness-of-fit tests. This points to a mechanism that amplifies an initial small perturbation, rather than to the sum of many random events.

\item The $m=1$ and $m=2$ modes are significantly coupled, with
$A_2 = 0.29\,A_1 + 0.11$ (Pearson $r = 0.27$, Spearman $\rho = 0.50$,
$p < 10^{-19}$). This is the largest-sample determination of the
$A_1$--$A_2$ relation to date.

\item $A_1$ increases monotonically with morphological T-type
($\tau = 0.23$, $p \sim 10^{-12}$), while $A_2$ shows no
significant morphological trend.

\item $A_1$ anti-correlates with the gas rotation velocity
($\rho = -0.432$, $p \sim 10^{-19}$) and, in the stellar-mass
cross-matched subsample of 246 galaxies with independent
$M^\star$ estimates, with $M^\star$ ($\rho = -0.395$, $p \sim 10^{-10}$).
The weaker lopsidedness of massive, fast-rotating disks indicates
that the observed $m=1$ amplitude is regulated by internal dynamics
rather than set by tidal perturbations alone.

\item The TNG50-1 simulation, analysed through the TNGGalaxyLab
Fourier pipeline, reproduces the observed $A_1$--$M^\star$
anti-correlation with a nearly identical Spearman coefficient
($\rho_S = -0.445$ against $-0.395$ observed), although its median
$A_1$ is $\sim 1.4\times$ lower than in the observations -- an offset
consistent with heterogeneous target selection in the observational
compilation.

\end{enumerate}

The compilation used in this work is publicly available through the
CDS/VizieR database as catalog VII/296 \citep{2025yCat.7296....0N}.
All figures and statistical results reported here can be reproduced
from the catalog using the analysis code publicly available at
\url{https://github.com/Abbos998/raa-lopsidedness-figures}.

\section*{Acknowledgements}

This research has made use of the VizieR catalogue access tool, CDS, Strasbourg, France (DOI: \url{10.26093/cds/vizier}), and the SIMBAD database, operated at CDS. This work was supported by the National University of Uzbekistan. We also acknowledge support from the Ministry of Higher Education, Science and Innovation of the Republic of Uzbekistan through Grant FZ-2020092851, as well as budgetary funding from the Academy of Sciences of the Republic of Uzbekistan.

\bibliographystyle{raa}
\bibliography{bibtex}

@ARTICLE{2000ApJ...542..216J,
       author = {{Jog}, Chanda J.},
        title = "{Self-consistent Response of a Galactic Disk to an Elliptical Perturbation Halo Potential}",
      journal = {\apj},
         year = 2000,
        month = oct,
       volume = {542},
       number = {1},
        pages = {216-223},
          doi = {10.1086/309521},
archivePrefix = {arXiv},
       eprint = {astro-ph/0007364},
 primaryClass = {astro-ph},
       adsurl = {https://ui.adsabs.harvard.edu/abs/2000ApJ...542..216J}
}

@ARTICLE{2009ApJ...691.1005R,
       author = {{Reichard}, Timothy A. and {Heckman}, Timothy M. and {Rudnick}, Gregory and {Brinchmann}, Jarle and {Kauffmann}, Guinevere and {Wild}, Vivienne},
        title = "{The Lopsidedness of Present-Day Galaxies: Connections to the Formation of Stars, the Chemical Evolution of Galaxies, and the Growth of Black Holes}",
      journal = {\apj},
         year = 2009,
        month = feb,
       volume = {691},
       number = {2},
        pages = {1005-1020},
          doi = {10.1088/0004-637X/691/2/1005},
archivePrefix = {arXiv},
       eprint = {0809.3310},
 primaryClass = {astro-ph},
       adsurl = {https://ui.adsabs.harvard.edu/abs/2009ApJ...691.1005R}
}

@INPROCEEDINGS{2001ASPC..230..457V,
       author = {{Vesperini}, E. and {Weinberg}, M.~D.},
        title = "{Perturbation of Dark Halos and Elliptical Galaxies During Flyby Encounters}",
    booktitle = {Galaxy Disks and Disk Galaxies},
         year = 2001,
       editor = {{Funes}, Jos{\'e} G. and {Corsini}, Enrico Maria},
       series = {Astronomical Society of the Pacific Conference Series},
       volume = {230},
        month = jan,
        pages = {457-458},
       adsurl = {https://ui.adsabs.harvard.edu/abs/2001ASPC..230..457V}
}

@ARTICLE{2006MNRAS.369.1849A,
       author = {{Angiras}, R.~A. and {Jog}, C.~J. and {Omar}, A. and {Dwarakanath}, K.~S.},
        title = "{Origin of disc lopsidedness in the Eridanus group of galaxies}",
      journal = {\mnras},
         year = 2006,
        month = jul,
       volume = {369},
       number = {4},
        pages = {1849-1857},
          doi = {10.1111/j.1365-2966.2006.10418.x},
archivePrefix = {arXiv},
       eprint = {astro-ph/0604120},
 primaryClass = {astro-ph},
       adsurl = {https://ui.adsabs.harvard.edu/abs/2006MNRAS.369.1849A}
}

@ARTICLE{2009A&A...495..491A,
       author = {{Aguerri}, J.~A.~L. and {M{\'e}ndez-Abreu}, J. and {Corsini}, E.~M.},
        title = "{The population of barred galaxies in the local universe. I. Detection and characterisation of bars}",
      journal = {\aap},
         year = 2009,
        month = feb,
       volume = {495},
       number = {2},
        pages = {491-504},
          doi = {10.1051/0004-6361:200810931},
archivePrefix = {arXiv},
       eprint = {0901.2346},
 primaryClass = {astro-ph.GA},
       adsurl = {https://ui.adsabs.harvard.edu/abs/2009A&A...495..491A}
}

@ARTICLE{1980MNRAS.193..313B,
       author = {{Baldwin}, J.~E. and {Lynden-Bell}, D. and {Sancisi}, R.},
        title = "{Lopsided galaxies}",
      journal = {\mnras},
         year = 1980,
        month = oct,
       volume = {193},
        pages = {313-319},
          doi = {10.1093/mnras/193.2.313},
       adsurl = {https://ui.adsabs.harvard.edu/abs/1980MNRAS.193..313B}
}

@ARTICLE{2005A&A...438..507B,
       author = {{Bournaud}, F. and {Combes}, F. and {Jog}, C.~J. and {Puerari}, I.},
        title = "{Lopsided spiral galaxies: evidence for gas accretion}",
      journal = {\aap},
         year = 2005,
        month = aug,
       volume = {438},
       number = {2},
        pages = {507-520},
          doi = {10.1051/0004-6361:20052631},
archivePrefix = {arXiv},
       eprint = {astro-ph/0503314},
 primaryClass = {astro-ph},
       adsurl = {https://ui.adsabs.harvard.edu/abs/2005A&A...438..507B}
}

@ARTICLE{2023MNRAS.526..567D,
       author = {{Dolfi}, Arianna and {G{\'o}mez}, Facundo A. and {Monachesi}, Antonela and {Varela-Lavin}, Silvio and {Tissera}, Patricia B. and {Sif{\'o}n}, Crist{\'o}bal and {Galaz}, Gaspar},
        title = "{Lopsidedness as a tracer of early galactic assembly history}",
      journal = {\mnras},
         year = 2023,
        month = nov,
       volume = {526},
       number = {1},
        pages = {567-584},
          doi = {10.1093/mnras/stad2650},
archivePrefix = {arXiv},
       eprint = {2306.04639},
 primaryClass = {astro-ph.GA},
       adsurl = {https://ui.adsabs.harvard.edu/abs/2023MNRAS.526..567D}
}

@ARTICLE{2025A&A...699A..11D,
       author = {{Dolfi}, Arianna and {G{\'o}mez}, Facundo A. and {Monachesi}, Antonela and {Tissera}, Patricia B. and {Sif{\'o}n}, Crist{\'o}bal and {Galaz}, Gaspar},
        title = "{Tracing the origins of galaxy lopsidedness across cosmic time}",
      journal = {\aap},
         year = 2025,
        month = jul,
       volume = {699},
          eid = {A11},
        pages = {A11},
          doi = {10.1051/0004-6361/202453216},
archivePrefix = {arXiv},
       eprint = {2411.19426},
 primaryClass = {astro-ph.GA},
       adsurl = {https://ui.adsabs.harvard.edu/abs/2025A&A...699A..11D}
}

@ARTICLE{1970ApJ...160..811F,
       author = {{Freeman}, K.~C.},
        title = "{On the Disks of Spiral and S0 Galaxies}",
      journal = {\apj},
         year = 1970,
        month = jun,
       volume = {160},
        pages = {811},
          doi = {10.1086/150474},
       adsurl = {https://ui.adsabs.harvard.edu/abs/1970ApJ...160..811F}
}

@ARTICLE{1998AJ....115...62H,
       author = {{Haynes}, Martha P. and {Hogg}, David E. and {Maddalena}, Ronald J. and {Roberts}, Morton S. and {van Zee}, Liese},
        title = "{Asymmetry in High-Precision Global H i Profiles of Isolated Spiral Galaxies}",
      journal = {\aj},
         year = 1998,
        month = jan,
       volume = {115},
       number = {1},
        pages = {62-79},
          doi = {10.1086/300166},
       adsurl = {https://ui.adsabs.harvard.edu/abs/1998AJ....115...62H}
}

@ARTICLE{2006MNRAS.370..891J,
       author = {{Jog}, Chanda J. and {Maybhate}, Aparna},
        title = "{Measurement of non-axisymmetry in centres of advanced mergers of galaxies}",
      journal = {\mnras},
         year = 2006,
        month = aug,
       volume = {370},
       number = {2},
        pages = {891-901},
          doi = {10.1111/j.1365-2966.2006.10529.x},
archivePrefix = {arXiv},
       eprint = {astro-ph/0605251},
 primaryClass = {astro-ph},
       adsurl = {https://ui.adsabs.harvard.edu/abs/2006MNRAS.370..891J}
}

@ARTICLE{2009PhR...471...75J,
       author = {{Jog}, Chanda J. and {Combes}, Francoise},
        title = "{Lopsided spiral galaxies}",
      journal = {\physrep},
         year = 2009,
        month = feb,
       volume = {471},
       number = {2},
        pages = {75-111},
          doi = {10.1016/j.physrep.2008.12.002},
archivePrefix = {arXiv},
       eprint = {0811.1101},
 primaryClass = {astro-ph},
       adsurl = {https://ui.adsabs.harvard.edu/abs/2009PhR...471...75J}
}

@ARTICLE{2022A&A...662A..53L,
       author = {{{\L}okas}, Ewa L.},
        title = "{Lopsided galactic disks in IllustrisTNG}",
      journal = {\aap},
         year = 2022,
        month = jun,
       volume = {662},
          eid = {A53},
        pages = {A53},
          doi = {10.1051/0004-6361/202142845},
archivePrefix = {arXiv},
       eprint = {2204.01456},
 primaryClass = {astro-ph.GA},
       adsurl = {https://ui.adsabs.harvard.edu/abs/2022A&A...662A..53L}
}

@ARTICLE{2019ComAC...6....2N,
       author = {{Nelson}, Dylan and {Springel}, Volker and {Pillepich}, Annalisa and {Rodriguez-Gomez}, Vicente and {Torrey}, Paul and {Genel}, Shy and {Vogelsberger}, Mark and {Pakmor}, Ruediger and {Marinacci}, Federico and {Weinberger}, Rainer and {Kelley}, Luke and {Lovell}, Mark and {Diemer}, Benedikt and {Hernquist}, Lars},
        title = "{The IllustrisTNG simulations: public data release}",
      journal = {Computational Astrophysics and Cosmology},
         year = 2019,
        month = may,
       volume = {6},
       number = {1},
          eid = {2},
        pages = {2},
          doi = {10.1186/s40668-019-0028-x},
archivePrefix = {arXiv},
       eprint = {1812.05609},
 primaryClass = {astro-ph.GA},
       adsurl = {https://ui.adsabs.harvard.edu/abs/2019ComAC...6....2N}
}

@ARTICLE{2011Ap.....54..184M,
       author = {{Mirtadjieva}, K.~T. and {Nuritdinov}, S.~N. and {Ruzibaev}, J.~K. and {Khalid}, Muhammad},
        title = "{Instabilities in a nonstationary model of self-gravitating disks. III. The phenomenon of lopsidedness and a comparison of perturbation modes}",
      journal = {Astrophysics},
         year = 2011,
        month = jun,
       volume = {54},
       number = {2},
        pages = {184-202},
          doi = {10.1007/s10511-011-9171-z},
       adsurl = {https://ui.adsabs.harvard.edu/abs/2011Ap.....54..184M}
}

@ARTICLE{2024AstBu..79..221M,
       author = {{Mirtadjieva}, K.~T. and {Nuritdinov}, S.~N. and {Omonov}, A.~U.},
        title = "{On the Theory of Formation of Lopsided Spiral Galaxies}",
      journal = {Astrophysical Bulletin},
         year = 2024,
        month = jun,
       volume = {79},
       number = {2},
        pages = {221-234},
          doi = {10.1134/S1990341324600017},
       adsurl = {https://ui.adsabs.harvard.edu/abs/2024AstBu..79..221M}
}

@dataset{2025yCat.7296....0N,
       author = {{Nuritdinov}, S.~N. and {Mirtadjieva}, K.~T. and {Omonov}, A.~U.},
        title = "{VizieR Online Data Catalog: Catalog of lopsided disk-like galaxies (Nuritdinov+, 2024)}",
 howpublished = {VizieR On-line Data Catalog: VII/296.  Originally published in: Modern astronomy, 1, 401-404 (2024)},
         year = 2025,
        month = jan,
          eid = {VII/296},
       adsurl = {https://ui.adsabs.harvard.edu/abs/2025yCat.7296....0N}
}

@ARTICLE{2019MNRAS.490.3196P,
       author = {{Pillepich}, Annalisa and {Nelson}, Dylan and {Springel}, Volker and {Pakmor}, R{\"u}diger and {Torrey}, Paul and {Weinberger}, Rainer and {Vogelsberger}, Mark and {Marinacci}, Federico and {Genel}, Shy and {van der Wel}, Arjen and {Hernquist}, Lars},
        title = "{First results from the TNG50 simulation: the evolution of stellar and gaseous discs across cosmic time}",
      journal = {\mnras},
         year = 2019,
        month = dec,
       volume = {490},
       number = {3},
        pages = {3196-3233},
          doi = {10.1093/mnras/stz2338},
archivePrefix = {arXiv},
       eprint = {1902.05553},
 primaryClass = {astro-ph.GA},
       adsurl = {https://ui.adsabs.harvard.edu/abs/2019MNRAS.490.3196P}
}

@ARTICLE{2003MNRAS.338...14P,
       author = {{Power}, C. and {Navarro}, J.~F. and {Jenkins}, A. and {Frenk}, C.~S. and {White}, S.~D.~M. and {Springel}, V. and {Stadel}, J. and {Quinn}, T.},
        title = "{The inner structure of {\ensuremath{\Lambda}}CDM haloes - I. A numerical convergence study}",
      journal = {\mnras},
         year = 2003,
        month = jan,
       volume = {338},
       number = {1},
        pages = {14-34},
          doi = {10.1046/j.1365-8711.2003.05925.x},
archivePrefix = {arXiv},
       eprint = {astro-ph/0201544},
 primaryClass = {astro-ph},
       adsurl = {https://ui.adsabs.harvard.edu/abs/2003MNRAS.338...14P}
}

@ARTICLE{1995ApJ...447...82R,
       author = {{Rix}, Hans-Walter and {Zaritsky}, Dennis},
        title = "{Nonaxisymmetric Structures in the Stellar Disks of Galaxies}",
      journal = {\apj},
         year = 1995,
        month = jul,
       volume = {447},
        pages = {82},
          doi = {10.1086/175858},
archivePrefix = {arXiv},
       eprint = {astro-ph/9505111},
 primaryClass = {astro-ph},
       adsurl = {https://ui.adsabs.harvard.edu/abs/1995ApJ...447...82R}
}

@ARTICLE{2009ApJ...697.2015S,
       author = {{Saha}, Kanak and {Levine}, Evan S. and {Jog}, Chanda J. and {Blitz}, Leo},
        title = "{The Milky Way's Dark Matter Halo Appears to be Lopsided}",
      journal = {\apj},
         year = 2009,
        month = jun,
       volume = {697},
       number = {2},
        pages = {2015-2029},
          doi = {10.1088/0004-637X/697/2/2015},
archivePrefix = {arXiv},
       eprint = {0903.3802},
 primaryClass = {astro-ph.GA},
       adsurl = {https://ui.adsabs.harvard.edu/abs/2009ApJ...697.2015S}
}

@ARTICLE{2010MNRAS.401..791S,
       author = {{Springel}, Volker},
        title = "{E pur si muove: Galilean-invariant cosmological hydrodynamical simulations on a moving mesh}",
      journal = {\mnras},
         year = 2010,
        month = jan,
       volume = {401},
       number = {2},
        pages = {791-851},
          doi = {10.1111/j.1365-2966.2009.15715.x},
archivePrefix = {arXiv},
       eprint = {0901.4107},
 primaryClass = {astro-ph.CO},
       adsurl = {https://ui.adsabs.harvard.edu/abs/2010MNRAS.401..791S}
}

@ARTICLE{1999MNRAS.304..330S,
       author = {{Swaters}, R.~A. and {Schoenmakers}, R.~H.~M. and {Sancisi}, R. and {van Albada}, T.~S.},
        title = "{Kinematically lopsided spiral galaxies}",
      journal = {\mnras},
         year = 1999,
        month = apr,
       volume = {304},
       number = {2},
        pages = {330-334},
          doi = {10.1046/j.1365-8711.1999.02332.x},
archivePrefix = {arXiv},
       eprint = {astro-ph/9811424},
 primaryClass = {astro-ph},
       adsurl = {https://ui.adsabs.harvard.edu/abs/1999MNRAS.304..330S}
}

@ARTICLE{2023MNRAS.523.5853V,
       author = {{Varela-Lavin}, Silvio and {G{\'o}mez}, Facundo A. and {Tissera}, Patricia B. and {Besla}, Gurtina and {Garavito-Camargo}, Nicol{\'a}s and {Marinacci}, Federico and {Laporte}, Chervin F.~P.},
        title = "{Lopsided galaxies in a cosmological context: a new galaxy-halo connection}",
      journal = {\mnras},
         year = 2023,
        month = aug,
       volume = {523},
       number = {4},
        pages = {5853-5868},
          doi = {10.1093/mnras/stad1724},
archivePrefix = {arXiv},
       eprint = {2211.16577},
 primaryClass = {astro-ph.GA},
       adsurl = {https://ui.adsabs.harvard.edu/abs/2023MNRAS.523.5853V}
}

@ARTICLE{2011A&A...530A..29V,
       author = {{van Eymeren}, J. and {J{\"u}tte}, E. and {Jog}, C.~J. and {Stein}, Y. and {Dettmar}, R.-J.},
        title = "{Lopsidedness in WHISP galaxies. I. Rotation curves and kinematic lopsidedness}",
      journal = {\aap},
         year = 2011,
        month = jun,
       volume = {530},
          eid = {A29},
        pages = {A29},
          doi = {10.1051/0004-6361/201016177},
archivePrefix = {arXiv},
       eprint = {1103.4928},
 primaryClass = {astro-ph.CO},
       adsurl = {https://ui.adsabs.harvard.edu/abs/2011A&A...530A..29V}
}

@ARTICLE{1995ApJ...455L..31W,
       author = {{Weinberg}, Martin D.},
        title = "{Production of Milky Way Structure by the Magellanic Clouds}",
      journal = {\apjl},
         year = 1995,
        month = dec,
       volume = {455},
        pages = {L31},
          doi = {10.1086/309803},
archivePrefix = {arXiv},
       eprint = {astro-ph/9507071},
 primaryClass = {astro-ph},
       adsurl = {https://ui.adsabs.harvard.edu/abs/1995ApJ...455L..31W}
}

@ARTICLE{1997ApJ...477..118Z,
       author = {{Zaritsky}, Dennis and {Rix}, Hans-Walter},
        title = "{Lopsided Spiral Galaxies and a Limit on the Galaxy Accretion Rate}",
      journal = {\apj},
         year = 1997,
        month = mar,
       volume = {477},
       number = {1},
        pages = {118-127},
          doi = {10.1086/303692},
archivePrefix = {arXiv},
       eprint = {astro-ph/9608086},
 primaryClass = {astro-ph},
       adsurl = {https://ui.adsabs.harvard.edu/abs/1997ApJ...477..118Z}
}

@ARTICLE{2013ApJ...772..135Z,
       author = {{Zaritsky}, Dennis and {Salo}, Heikki and {Laurikainen}, Eija and {Elmegreen}, Debra and {Athanassoula}, E. and {Bosma}, Albert and {Comer{\'o}n}, S{\'e}bastien and {Erroz-Ferrer}, Santiago and {Elmegreen}, Bruce and {Gadotti}, Dimitri A. and {Gil de Paz}, Armando and {Hinz}, Joannah L. and {Ho}, Luis C. and {Holwerda}, Benne W. and {Kim}, Taehyun and {Knapen}, Johan H. and {Laine}, Jarkko and {Laine}, Seppo and {Madore}, Barry F. and {Meidt}, Sharon and {Menendez-Delmestre}, Karin and {Mizusawa}, Trisha and {Mu{\~n}oz-Mateos}, Juan Carlos and {Regan}, Michael W. and {Seibert}, Mark and {Sheth}, Kartik},
        title = "{On the Origin of Lopsidedness in Galaxies as Determined from the Spitzer Survey of Stellar Structure in Galaxies (S$^{4}$G)}",
      journal = {\apj},
         year = 2013,
        month = aug,
       volume = {772},
       number = {2},
          eid = {135},
        pages = {135},
          doi = {10.1088/0004-637X/772/2/135},
archivePrefix = {arXiv},
       eprint = {1305.2940},
 primaryClass = {astro-ph.CO},
       adsurl = {https://ui.adsabs.harvard.edu/abs/2013ApJ...772..135Z}
}

\end{document}